%% file: main.tex
\documentclass[conference]{IEEEtran}
\usepackage{cite}
\usepackage{amsmath,amssymb,amsfonts}
\usepackage{graphicx}
\usepackage{booktabs}
\usepackage{array}
\usepackage{multirow}
\usepackage[hyphens]{url}
\usepackage{xcolor}
\usepackage{microtype}
\usepackage{eso-pic}
\begin{document}
\AddToShipoutPictureFG*{\AtPageLowerLeft{\put(\LenToUnit{0.6in},\LenToUnit{0.22in}){\parbox[b]{\dimexpr\paperwidth-1.2in\relax}{\fontsize{6.5}{7.5}\selectfont\centering \copyright\ 2026 IEEE. Personal use of this material is permitted. Permission from IEEE must be obtained for all other uses, in any current or future media, including reprinting/republishing this material for advertising or promotional purposes, creating new collective works, for resale or redistribution to servers or lists, or reuse of any copyrighted component of this work in other works. Accepted to the 2026 IEEE International Workshop on Information Forensics and Security (WIFS).}}}}
\pdfinfo{/Title (Detecting Generated-Depth Injection Attacks on RGB-D Place Authentication) /Author (Seigo Kawachi and Satoshi Matsuura)}
\title{Detecting Generated-Depth Injection Attacks on RGB-D Place Authentication}
\author{\IEEEauthorblockN{Seigo Kawachi\IEEEauthorrefmark{1} and Satoshi Matsuura\IEEEauthorrefmark{1}}
\IEEEauthorblockA{\IEEEauthorrefmark{1}\textit{Dept.\ of Mathematical and Computing Science, School of Computing} \\
\textit{Institute of Science Tokyo}, Tokyo, Japan \\ kawachi.s.62d3@m.isct.ac.jp}}
\maketitle
\begin{abstract}\input{sections/00_abstract}\end{abstract}
\input{sections/01_intro}
\input{sections/02_threat}
\input{sections/03_method}
\input{sections/04_indomain}
\input{sections/05_boundaries}
\input{sections/06_discussion}
\section*{Acknowledgment}
This work was supported in part by Palo Alto Networks. The authors used Claude (Anthropic) to help cross-check numerical results against the experimental records, make minor language edits, and prepare the LaTeX manuscript. All outputs produced with AI assistance were reviewed and verified by the authors.
\bibliographystyle{IEEEtran}
\bibliography{refs}
\end{document}

%% file: sections/00_abstract.tex
Advances in monocular depth estimation (MDE) now let an attacker generate depth from publicly available RGB images and submit forged RGB-D queries. Such attacks break a core assumption of RGB-D place authentication: that depth comes from a sensor. We study a presentation attack detection (PAD) classifier that uses geometric consistency between the enrollment and query views, as a software-layer defense. A three-stage evaluation shows how far this classifier detects software-injected depth, depending on how accurately the attacker knows the camera intrinsics. In the first stage, the attacker has only an estimate of the intrinsics, and the PAD classifier separates forgeries generated by all three MDE models with a validation area under the receiver operating characteristic curve (AUC) of 0.997--0.999. In the second stage, the attacker obtains the exact intrinsics, and the original PAD classifier falls to near chance (AUC 0.54--0.61); a retrained PAD classifier, however, still has the registration residual between enrollment and query as a cue. In the third stage, an oracle evaluation, in which the evaluator fits warps using the ground-truth 3D geometry, shows that detection falls to near chance only in its strongest case, where every valid query keypoint is given its ground-truth 3D coordinate. Of the detection cues, the reprojection error captures the attacker's misestimate of the intrinsics, whereas the registration residual remains even when the attacker knows the exact intrinsics.

%% file: sections/01_intro.tex
\section{Introduction}\label{sec:intro}

Place authentication verifies that a device is at a previously enrolled location using the geometry of that environment \cite{worldauth,worldauth_lidar}. Platforms such as Apple ARKit and Google ARCore Raw Depth have made place authentication practical on devices that measure depth with LiDAR or ToF sensors. Compared with appearance-only matching, depth imposes a physical constraint on the attacker. In principle, producing a geometrically consistent RGB-depth pair requires the attacker to be at the target location. An attacker who fools this verification can bypass location-based access control without being physically present.

Recent monocular depth estimation (MDE) models weaken this physical constraint by allowing an attacker to generate depth from public RGB images alone. An attacker can use Depth Anything 3 (DA3) \cite{da3}, VGGT \cite{vggt}, or Marigold \cite{marigold} to construct forged RGB-D queries that bypass the assumption of sensor-acquired depth. Place authentication still retains one cue that single-acquisition forgery detectors do not have. It matches local features between an enrollment view and a query view, and these correspondences make the geometric consistency between the two views available for detection. We detect injected depth with a presentation attack detection (PAD) classifier that uses this enrollment--query geometric consistency.

Because the device back-projects keypoints with the camera intrinsics $K$ (focal length and principal point), how accurately the attacker knows $K$ governs detection. Following Kerckhoffs's principle, we assume that the attacker knows the verification algorithm, the feature extraction procedure, the random forest (RF) classifiers (authentication and PAD), and their training data, and we do not rely on secrecy of $K$. We vary how accurately the attacker knows $K$, from an estimate to the true value.

The evaluation asks two questions. First, which geometric cues remain for the PAD classifier as the attacker learns more about $K$? Second, can the PAD classifier still reject a forged query whose viewpoint already lies inside the authentication acceptance region defined by pose?

\begin{enumerate}
\item \textbf{Stage 1 -- mismatched-$K$ (Section~\ref{sec:indomain}):} The attacker estimates $K$. On Washington RGB-D Scenes v2 \cite{washington}, the PAD classifier separates forgeries with near-perfect validation area under the receiver operating characteristic curve (AUC). This separation comes mainly from the mismatch between the attacker's $K$ and the verifier's $K$, not from the generated depth itself.
\item \textbf{Stage 2 -- exact-$K$ (Section~\ref{sec:exactk}):} The attacker knows the verifier's exact $K$. The mismatch disappears, and the original PAD classifier falls to near chance. A retrained PAD classifier first relies on the keypoint-rounding difference between genuine and forged queries; once the attacker also matches the keypoint rounding, the registration residual, a cross-view cue, remains as the main cue.
\item \textbf{Stage 3 -- acceptance region (Section~\ref{sec:inband}):} The attacker uses the same $K$ as the verifier, so that neither back-projection nor reprojection carries a $K$ error, and the query lies inside the acceptance region. In an oracle evaluation, the evaluator fits warps using the ground-truth 3D geometry. Among the evaluated warps, detection falls to near chance only when every valid query keypoint is given its ground-truth 3D coordinate.
\end{enumerate}

The central security finding of this paper is the decomposition of detection into a part that depends on the $K$ mismatch and a part that remains when the attacker knows $K$. The former is lost under exact-$K$, where the original PAD classifier falls to near chance. In the latter, the registration residual remains.

Earlier studies on place authentication, including geometry-based methods \cite{worldauth,worldauth_lidar} and a point-cloud method based on intrinsic shape signature keypoints evaluated on ARKitScenes \cite{issregauth}, assume genuine acquisition and do not study an attacker who controls the submitted depth. We address that gap with a threat model in which software injects generated depth and the verifier evaluates the resulting query by its geometric consistency with the enrollment.

The closest studies on depth forgery operate on a single acquisition. DepthFake \cite{wu2023depthfake} estimates facial depth from a single 2D photo to spoof 3D face authentication. Scoop \cite{scoop} detects recapture attacks by comparing learned depth with the measured time-of-flight (ToF) depth; it explicitly excludes both sensor-bypass attacks and two-view cues from its scope. Chiesa et al. \cite{chiesa2021gan} injected GAN-generated depth to attack single-acquisition PAD for RGB-D face recognition. Our work complements these studies with the sensor-bypass, two-view setting, in which the attacker injects generated depth after the sensor.

%% file: sections/02_threat.tex
\section{Threat Model and System}\label{sec:threat}

\begin{figure*}[t]
\centering
\includegraphics[width=\textwidth]{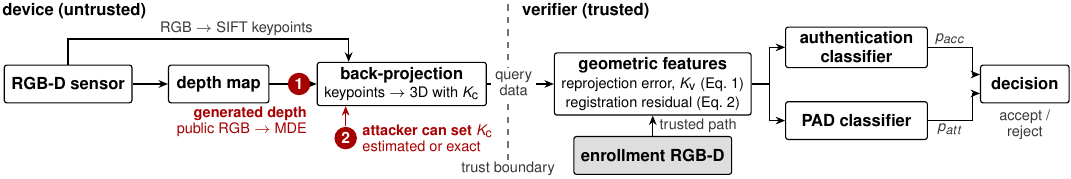}
\caption{System and threat model. Red numbers mark the attack points:
(1) depth generated from public RGB replaces the sensor depth before
back-projection, and (2) the attacker can set the intrinsics $K_{\mathrm{c}}$.
Only query keypoints, descriptors, and 3D points cross the trust boundary.}
\label{fig:threat}
\end{figure*}

At authentication time, the verifier does not receive a depth image. It receives Scale-Invariant Feature Transform (SIFT) keypoints and their descriptors, together with 3D points that the device back-projects from the keypoints using the depth map and its own $K$. The verifier reprojects these points using its own $K$ (Fig.~\ref{fig:threat}). The enrollment RGB-D reaches the verifier over a trusted path. The verifier uses the same geometric features for two separate classifiers. The authentication classifier, with output $p_{acc}$, determines whether the query comes from the same place. The PAD classifier, with output $p_{att}$, determines whether the submitted depth is forged.

The authentication acceptance region (hereafter, the acceptance region) is the range of viewpoints for which the authentication classifier is trained to accept a query: $\Delta\mathrm{pos}\le0.20$ m and $\Delta\theta\le12^{\circ}$. The authentication classifier is trained to reject queries outside it as place mismatches, and the PAD classifier evaluates queries inside it. When the two decisions are combined, the system accepts only if $p_{acc}\ge\tau_{\mathrm{auth}}$ and $p_{att}<\tau_{\mathrm{pad}}$, where $\tau_{\mathrm{auth}}$ and $\tau_{\mathrm{pad}}$ are the authentication and PAD thresholds.

The attack injects depth after sensor acquisition but before device-side back-projection. A tamper-resistant sensor alone does not protect this downstream software path. On consumer devices where hardware does not protect the entire path from sensor output to the back-projection input, the detector acts as a software-layer defense. If that entire path is hardware-protected, the injection considered here is not possible. We assume that genuine queries back-project keypoints using sensor-acquired depth from LiDAR, structured light, or ToF. Configurations that legitimately use estimated depth in the normal path, such as ARCore on devices without ToF, are discussed in Section~\ref{sec:exactk} and remain outside the main threat model.

Three attacks lie outside our scope. Freshness checks and nonces address replay, a trusted enrollment process such as identity verification or device attestation prevents enrollment compromise, and an attacker with physical access falls outside the remote threat model considered here.

Besides the knowledge assumed in Section~\ref{sec:intro}, the attacker has public RGB imagery of the target location and an MDE model, but lacks physical access, an on-site depth sensor, and enrollment data. Public RGB may come from social media or Street View imagery.

Marigold outputs affine-invariant relative depth. The attacker converts it to metric depth using the fixed heuristic

$\mathrm{depth}_{\mathrm{metric}}=0.4+2.6\cdot\mathrm{depth}_{\mathrm{rel}}$,

based on the typical Washington depth range. This prior is not fitted to any training, validation, or test split. It gives the attacker prior information about scene depth and therefore makes the evaluation more favorable to the attacker.

We evaluate two acquisition conditions. Under ideal acquisition, the attacker has the query image and its SIFT keypoints and forges only the depth. Genuine and attack pairs then have bit-identical RGB, keypoints, and pose for any choice of $K$, so any separation stems only from the submitted depth and the resulting geometry. Under off-viewpoint acquisition, the attacker has only public RGB from a different viewpoint.

Because $K$ can be obtained from an augmented reality session or dataset calibration, we vary how accurately the attacker knows it. Under \emph{mismatched-$K$}, the attacker estimates $K$ from typical values. Under \emph{exact-$K$}, the attacker knows the verifier's true $K$. A stronger exact-$K$ attacker also matches the keypoint rounding of the genuine pipeline, which converts keypoint coordinates to integer pixels before back-projection; this implementation detail is observable from published code. Table~\ref{tab:main} summarizes the evaluated settings.

%% file: sections/03_method.tex
\section{Detector and Evaluation Protocol}\label{sec:method}

The PAD classifier uses five groups of geometric features computed from enrollment--query correspondences obtained with SIFT and Random Sample Consensus (RANSAC) \cite{lowe2004sift,fischler1981ransac}. The first group describes 2D matching. The second describes RANSAC geometry in terms of the number and ratio of inliers. The third contains statistics of the registration residual (Eq.~\ref{eq:residual}). The fourth contains depth statistics, including agreement between the normals of two dominant planes fitted by principal component analysis to the RANSAC inliers. The fifth measures consistency between depth and the camera model by aggregating per-frame reprojection error (Eq.~\ref{eq:reprojection}) and its inlier ratio for the enrollment and query frames.

Plane filtering produces the final inlier set used by the reprojection features. The registration residual uses the RANSAC inliers before this filtering step. Undefined features receive a fixed sentinel value.

The reprojection error measures within-frame consistency between device-side back-projection and verifier-side reprojection (Eq.~\ref{eq:reprojection}). For each final inlier, it is the positional offset between the original keypoint and the same 3D point projected back into the image. Because the computation stays within one frame, it measures disagreement between the device-side $K$ and the verifier-side $K$. This feature separates Stage 1 from Stage 2.

\begin{equation}
e_i = \bigl\| \pi_{K_{\mathrm{v}}}(X_i) - u_i \bigr\|_2,
\qquad
X_i = d_i\, K_{\mathrm{c}}^{-1} \bigl[u_i^{\top},\,1\bigr]^{\top},
\label{eq:reprojection}
\end{equation}
where $u_i$ is the $i$-th SIFT keypoint of a frame, $d_i$ its depth,
$K_{\mathrm{c}}$ the intrinsics used for device-side back-projection,
$K_{\mathrm{v}}$ the verifier's fixed intrinsics, and
$\pi_K(\cdot)$ the pinhole projection with $K$; genuine queries round $u_i$ to integer pixels in $X_i$.

The registration residual provides the complementary cross-view cue (Eq.~\ref{eq:residual}). It is the per-correspondence 3D distance that remains after a rigid transform maps enrollment-side inlier points into the query coordinate frame. This quantity does not involve the within-frame $K$ round trip, and it is the main cue left after retraining under exact-$K$ (Section~\ref{sec:exactk}).

\begin{equation}
\begin{aligned}
r_j &= \bigl\| R\,P_j + t - Q_j \bigr\|_2,\\
(R,t) &= \operatorname*{arg\,min}_{R \in SO(3),\, t} \sum_j \bigl\| R\,P_j + t - Q_j \bigr\|_2^2,
\end{aligned}
\label{eq:residual}
\end{equation}
where $(P_j, Q_j)$ is the $j$-th RANSAC-inlier correspondence between the
enrollment and query 3D points. The registration-residual features are the mean,
median, and standard deviation of $r_j$ over these inliers.

The full feature extractor produces 49 geometric features. We remove six features that reflect the acquisition process or dataset instead of scene geometry, so the main PAD experiments use 43 features; the W0--W4 evaluation in Section~\ref{sec:inband} uses a 37-feature set from the on-device implementation, which shares 35 features with the main set.

The PAD classifier is an RF trained on the 43-feature set. It uses 200 trees, maximum depth 10, balanced class weights, and random seed 42. The authentication classifier is trained separately on all available features, with labels given by the acceptance region of Section~\ref{sec:threat}.

The implementation uses OpenCV 4.11 for SIFT and RANSAC and scikit-learn 1.8 under Python 3.12. The device implementation uses OpenCV 4.10 and ONNX Runtime 1.18 on CPU.

We evaluate on Washington RGB-D Scenes v2 \cite{washington} and ARKitScenes \cite{arkitscenes}. For Washington, scenes 1--8 form the training set, scenes 9--11 the validation set, and scenes 12--14 the held-out test set. The held-out test set is never used for model or threshold selection. For ARKitScenes, we define a visit group as all samples sharing the same visit ID and use a deterministic 70/30 split by visit group, keeping each visit group entirely within one partition.

We report PAD performance with AUC plus the operating-point metrics of ISO/IEC 30107-3 \cite{iso30107_3}: the attack presentation classification error rate (APCER) at a fixed bona fide presentation classification error rate (BPCER). Depth injected after the sensor is an injection attack in the terminology of CEN/TS 18099 \cite{cents18099}; for comparability with prior PAD work, we use the PAD, APCER, and BPCER terminology and refer to bona fide samples as genuine. Per-model results are reported in the order DA3 / VGGT / Marigold.

We measure uncertainty with bootstrap confidence intervals (CIs, $B=1000$), stating the resampling unit for each analysis, and, where applicable, with variation across 10 RF seeds. Ablations are additionally checked with paired DeLong tests under Holm--Bonferroni correction and with label-permutation null distributions ($B=1000$). Cross-dataset splits follow the grouping principle used in face PAD \cite{costapazo}, grouping samples by visit instead of using strict leave-one-out.

To identify which feature families drive separation, we evaluate five detector variants. \emph{Per-frame quality and descriptor statistics} uses only depth-related quality and descriptor features and approximates single-acquisition depth-forgery detectors. \emph{Geometry without depth statistics} removes the depth-statistics features. \emph{Without reprojection features} removes only the reprojection-error and reprojection-inlier features and keeps the registration-residual and RANSAC-geometry features. \emph{Without two-view geometry features} removes the reprojection and registration-residual features and most RANSAC-geometry features, keeping 2D matching, plane agreement, and two RANSAC inlier ratios. Finally, \emph{logistic regression} replaces the RF while using the same 43-feature input.

The full PAD pipeline takes a median of 474 ms per pair on the CPU of a Redmi Note 11 consumer Android phone with 640$\times$480 images and $n=50$ pairs; the timing includes input/output handling and all PAD stages, and the upper end of the 95\% CI is 483 ms. SIFT extraction from the two frames accounts for 78\% of the median runtime, geometric-feature computation for 13\%, back-projection plus keypoint matching and RANSAC for 8\%, and RF inference for less than 1\%.

%% file: sections/04_indomain.tex
\section{Stage 1: Mismatched-$K$}\label{sec:indomain}

\begin{table}[t]
\centering
\caption{PAD AUC by Attack Stage}
\label{tab:main}
\scriptsize
\setlength{\tabcolsep}{3pt}
\begin{tabular*}{\columnwidth}{@{\extracolsep{\fill}}ll>{\centering\arraybackslash}p{.11\columnwidth}>{\centering\arraybackslash}p{.11\columnwidth}>{\centering\arraybackslash}p{.11\columnwidth}@{}}
\toprule
Attack & PAD classifier & DA3 & VGGT & Marigold \\
\midrule
\multicolumn{5}{@{}l}{\emph{Stage 1 (\S\ref{sec:indomain}): estimated $K$; near-perfect separation}} \\
Mismatched-$K$ & original & 0.997 & 0.999 & 0.997 \\
Mismatched-$K$ (test) & original & 0.993 & 0.996 & -- \\
\midrule
\multicolumn{5}{@{}l}{\emph{Stage 2 (\S\ref{sec:exactk}): verifier's $K$; original PAD near chance}} \\
Exact-$K$ & original & 0.542 & 0.613 & 0.577 \\
Exact-$K$ $+$ rounding & retrained, exact-$K$ & 0.478 & 0.479 & -- \\
Exact-$K$ $+$ rounding & retrained, rounding$^{*}$ & 0.704 & 0.817 & -- \\
\midrule
\multicolumn{5}{@{}l}{\emph{Stage 3 (\S\ref{sec:inband}): acceptance region}} \\
Off-viewpoint & \multicolumn{4}{l}{rejected by authentication RF: 0.0\% success at 1\% FPR} \\
Stress test & retrained, stress test & 0.992 & 0.980 & 0.999 \\
\midrule
\multicolumn{5}{@{}l}{\emph{Stage 3 oracle on test scenes, not realizable (\S\ref{sec:inband}): near chance only at W4}} \\
W0--W3 & retrained per stage & 1.000 & 1.000 & 1.000 \\
W4 & retrained per stage & 0.525 & 0.511 & 0.520 \\
\bottomrule
\end{tabular*}

\vspace{2pt}
\parbox{\columnwidth}{\scriptsize Washington validation scenes 9--11 unless marked test (scenes 12--14); ``--'': not measured. Exact-$K$, original: mean of 10 RF seeds (SD 0.10--0.17); $^{*}$3-fold out-of-fold. Stages 1--2 and the stress test assume ideal acquisition.}
\end{table}

Under mismatched-$K$, the PAD classifier separates forgeries with validation AUCs of 0.997 / 0.999 / 0.997. Changing the attacker's estimate of $K$ leaves the MDE depth output unchanged and affects only back-projection and downstream 3D features. The verifier reprojects those points with its own fixed $K$, producing a characteristic reprojection-error pattern. The importance analysis below attributes the separation to the intrinsics mismatch. At the operating point fixed at BPCER = 1\%, validation APCER is 0.058 / 0.028 for DA3 / VGGT.

A small intrinsics mismatch is enough for strong detection. With the PAD classifier fixed, we jointly vary the attacker's estimated focal length and principal point from +0.5\% to +10\% and evaluate on the validation set. AUC is about 0.92 at +0.5\%, rises monotonically for both DA3 and VGGT, reaches 0.997--0.999 at +2\%, and varies by less than 0.001 from +2\% to +10\%.

Detection under mismatched-$K$ transfers across the three evaluated MDE models, which cover feed-forward (DA3), multi-view (VGGT), and diffusion-based (Marigold) approaches. Across all $3\times3$ train/test combinations, within-model and cross-model AUCs are both approximately 0.9975, with a difference of only $\Delta=1.7\times10^{-4}$. The worst cross-model case is DA3$\rightarrow$Marigold at 0.9968.

Because the scenes do not overlap across splits, spatial overlap cannot explain the separation, which is statistically significant. In 1,000 label permutations on the 49-feature set, no permuted run reached the observed AUC ($p \le 0.001$). The same result was obtained with 200 label permutations on the 43-feature set for DA3 ($p \le 0.005$).

Reprojection features account for most of the separation under mismatched-$K$. All top 10 PAD-classifier features by Gini importance are statistics of reprojection error or reprojection inliers. The variant without two-view geometry features (Section~\ref{sec:method}) lowers AUC to 0.69--0.79, and the per-frame quality/descriptor variant reaches only 0.52; both differ significantly from the full 43-feature PAD classifier (paired DeLong tests, Holm-adjusted $p<0.001$). In contrast, removing only the depth-statistics features or replacing the RF with logistic regression makes no significant difference (Holm-adjusted $p>0.5$).

On the held-out Washington test, the PAD classifier trained on the training set loses at most 0.004 in AUC without further tuning. With the validation-fixed threshold, the operating point shifts more than the AUC: realized BPCER is 0.028 / 0.092 and APCER is 0.072 / 0.020 for DA3 / VGGT.

ARKitScenes provides a second in-domain test of the same feature pipeline. From a fixed split of 1168 training and 501 test visit groups, we subsample 200 training and 150 test groups. The aggregate AUC is 0.926 (95\% CI [0.916, 0.936], block bootstrap by visit group), and the per-model AUCs are 0.938 / 0.940 / 0.923. Without reprojection features, AUC falls to 0.60--0.70.

Part of the discriminative power of the reprojection features comes from differences in the acquisition path. A capture-condition control on ARKitScenes uses real sensor depth from a different capture session as genuine queries: enrollment uses frames from session A, and genuine queries use frames from session B of the same building. Forged queries use MDE depth generated from the same session-B frames and are reprojected with the verifier's $K$. Pairs are limited to the acceptance region, and the reprojection features are removed while the registration residual is kept (RF seed 42). Under this control, AUC stays at 0.92 for Marigold and is moderate for DA3 and VGGT (0.80 / 0.81); the 95\% CI lower bounds are 0.59 / 0.61 / 0.74 for DA3 / VGGT / Marigold. This control is a protocol separate from the in-domain evaluation above, not that evaluation with a control added. No control with genuine samples from a different capture session was run on Washington. The Washington AUC of 0.997--0.999 should therefore be read as a value under conditions that include the acquisition path of the evaluated pipeline, not as evidence of general robustness against generated depth.

Detection under mismatched-$K$ also survives across datasets: training on ARKitScenes and testing on Washington gives an AUC of at least 0.997. Transfer in the other direction is partial and asymmetric: training on Washington and testing on ARKitScenes gives AUCs of 0.887 / 0.860 / 0.868. Detection also holds against a limited adaptive attack that, unlike attacks optimizing through a surrogate model \cite{surrogate}, applies the Covariance Matrix Adaptation Evolution Strategy (CMA-ES) \cite{hansen2001cmaes} directly to the PAD score and varies five depth post-processing parameters: global scale, global offset, local Gaussian noise, near clipping, and far clipping. Across 600 validation pairs each for DA3 and VGGT, it lowers AUC by about 0.01. We do not evaluate this adaptive search for Marigold or under exact-$K$.

%% file: sections/05_boundaries.tex
\section{Exact-$K$ and Acceptance-Region Analyses}\label{sec:boundaries}

\subsection{Stage 2: Exact-$K$}\label{sec:exactk}

Under exact-$K$, reprojection error no longer separates genuine and forged queries. With the verifier's true $K$, forged queries have a measured reprojection error of approximately $5\times10^{-6}$ px. The remaining difference comes from the keypoint rounding defined in Section~\ref{sec:threat}: genuine queries retain a median reprojection error of about 0.4 px, whereas forged queries back-projected from floating-point coordinates do not. Under mismatched-$K$, the 24--28 px error caused by the intrinsics mismatch hides this 0.4 px difference.

When the attacker knows the exact $K$, the original PAD classifier falls to near chance for all three MDEs (Table~\ref{tab:main}; mean over 10 RF seeds). Extending the Section~\ref{sec:indomain} sweep to 0\% (DA3 and VGGT) gives AUCs of 0.561 / 0.638, within seed-to-seed variability.

The genuine and forged distributions remain partially separable under exact-$K$, so retraining can recover some signal. A PAD classifier retrained on exact-$K$ attacks initially relies mainly on the rounding difference between genuine and forged queries. When the attacker also matches the keypoint rounding, its AUC falls to 0.478 / 0.479 for DA3 / VGGT. Retraining again on this stronger attack yields validation AUCs (3-fold out-of-fold (OOF), one validation scene per fold) of 0.704 / 0.817 for DA3 / VGGT (Marigold was not measured). A PAD classifier without reprojection features, retrained on exact-$K$ attacks without rounding matching, reaches a similar level; because removing these features also removes the rounding cue, the remaining signal comes mainly from the registration residual.

Both the strong mismatched-$K$ result and its collapse under exact-$K$ depend on the system design in Section~\ref{sec:threat}, where the device sends locally back-projected 3D points. If the verifier receives a depth image and performs back-projection with its own $K$, the attacker's $K$ does not enter the round trip, and reprojection error is near zero for all queries. A same-$K$ pilot using the ARCore Depth API on a device without ToF ($n=50$) shows the same pattern as exact-$K$: reprojection error is near zero for both genuine and forged samples, and the registration residual is the main geometric difference.

\subsection{Stage 3: Authentication Acceptance Region}\label{sec:inband}

Stage 3 asks whether the PAD classifier can reject a forgery whose viewpoint already lies inside the acceptance region defined by pose.

A PAD classifier retrained on stress-test attacks detects forged queries whose viewpoints lie inside the acceptance region (validation AUC 0.980--0.999; Table~\ref{tab:main}). At BPCER 5\%, APCER is 0.047 / 0.087 / 0.000 for DA3 / VGGT / Marigold. This PAD classifier is trained on 11614 in-region pairs in total from the training scenes and evaluated on 2101 in-region pairs per class from the validation scenes; both training and evaluation use only pairs inside the acceptance region. The corresponding Stage 2 value is not the original PAD classifier's AUC of 0.54--0.61 but the 0.704 / 0.817 (DA3 / VGGT, 3-fold OOF, Section~\ref{sec:exactk}) of the PAD classifier retrained on rounding-matched attacks. Forged queries, genuine queries, and enrollment 3D points are all back-projected with the same $K$ as the verifier, and the verifier reprojects them with that $K$, so neither step carries a $K$ error and no keypoint-rounding cue exists. Only the query depth is forged: query SIFT keypoints are back-projected from generated depth, and the enrollment keeps genuine depth. For each pair, the attacker rescales the generated depth so that its median equals the enrollment median depth and marks depth as missing wherever the enrollment depth is missing. The threat model of Section~\ref{sec:threat} denies the attacker these enrollment statistics; the stress test deliberately grants them, and detection inside the acceptance region still holds under this attacker-favorable condition. Genuine and forged samples share the same pose distribution, so pose alone cannot explain the separation. Removing features that reflect capture confounds, that is, differences in capture session or device pipeline, leaves AUC unchanged ($\Delta=0.000$).

The authentication classifier rejects the executed DUSt3R-based \cite{dust3r} off-viewpoint reconstruction as a place mismatch. Unlike the oracle evaluations below, this attack was actually carried out. At the operating point where the authentication false-positive rate is 1\%, its attack success is 0.0\% on both the Washington validation and held-out test sets.

Next, under the same $K$ condition as the stress test, we apply a sequence of increasingly expressive warps, W0--W4. In this oracle evaluation, the evaluator, not the attacker, fits W0--W3 to the ground-truth query geometry, that is, the sensor-acquired 3D points of the genuine query frame, and chains each stage from the previous output. W0 is a similarity transform (7 parameters), W1 an affine transform (12 parameters), W2 a per-coordinate cubic polynomial (30 parameters), and W3 a dense grid-based warp (nominally 768 parameters). W4 removes the parameterization and replaces every valid query SIFT keypoint with its ground-truth 3D coordinate.

We evaluate W0--W4 on 2131 pairs from the held-out Washington test scenes (scenes 12--14) that lie inside the acceptance region. For each stage, we retrain the PAD classifier on the combined training and validation data. We also report a residual-only score that involves no classifier training. It is the median per-correspondence 3D distance after fitting a least-squares similarity transform to the RANSAC-inlier correspondences between the enrollment 3D points and the warped query 3D points. The correspondences and the distance are those of the registration residual in Eq.~\ref{eq:residual}; only the transform differs.

Both the PAD classifier retrained at each stage and the residual-only score separate forgeries through W3 (residual-only AUC 0.930--0.995) and fall to near chance at W4 (retrained PAD AUC 0.511--0.525, Table~\ref{tab:main}; residual-only AUC 0.513--0.518).

The same boundary appears without stage-specific retraining. For a PAD classifier trained only on W0, AUC remains at 1.000 through W3 and falls to 0.503--0.504 at W4. From W0 to W3, the fitted warps do not remove the local shape errors of the forged samples. The median warp residual, defined in the same way as the residual-only score, is 15.9--34.0 mm for forged samples and about 5.7 mm for genuine samples.

%% file: sections/06_discussion.tex
\section{Discussion, Limitations, and Conclusion}\label{sec:discussion}

The three stages show which geometric-consistency cues carry PAD detection, and which remain, as the attacker knows $K$ more accurately. When the attacker misestimates the camera intrinsics, the strongest signal comes from the reprojection error. When the attacker knows $K$, the original PAD classifier falls to near chance. However, a retrained PAD classifier retains the registration residual as a cue. It reaches validation AUCs of 0.70 / 0.82 (DA3 / VGGT) against attacks that also match keypoint rounding. Even when the attacker uses the same $K$ as the verifier, a PAD classifier retrained on stress-test attacks detects the stress-test forged queries inside the pose-defined acceptance region with AUC 0.980--0.999 when trained and evaluated only on pairs inside that region. In the oracle evaluation, where the evaluator fits warps using the ground-truth 3D geometry, detection holds through the dense warp W3 and falls to near chance only at W4, which gives every valid query SIFT keypoint its ground-truth 3D coordinate.

The major limitation of the method is that the original PAD classifier does not work as a defense when the attacker knows $K$ (Section~\ref{sec:exactk}). Against an attacker who also matches keypoint rounding, even a retrained PAD classifier separates forgeries only partially, relying mainly on the registration residual; this classifier should also be evaluated on the held-out test scenes. We evaluate authentication and PAD separately and do not report end-to-end false acceptance or false rejection rates for the combined system; such an evaluation requires matched pose-offset distributions for genuine and forged queries. Our experiments identify which geometric cues disappear and which remain as the attacker's knowledge grows; a fair comparison with detectors that learn two-view geometric consistency from the same inputs and threat model remains future work. We do not evaluate neural radiance field or 3D Gaussian Splatting reconstructions, unseen MDEs, broader adaptive attackers that explicitly account for $K$, or attackers who obtain the full enrollment depth. ARKitScenes contains indoor iPad LiDAR captures close to mobile-device use, and Washington contains indoor captures from a Kinect-type sensor. The results therefore do not establish performance on unseen devices, outdoors, or in substantially different sensing environments.

We studied attacks that inject depth after sensor acquisition and presented a PAD classifier that detects such forgeries through the geometric consistency between enrollment and query. When the attacker misestimated the camera intrinsics, the classifier detected forgeries generated by all three evaluated MDEs (validation AUC 0.997--0.999). When the attacker knew the exact intrinsics, the reprojection cue disappeared, but the registration residual remained as a cue. In the oracle evaluation, where the evaluator fitted warps using the ground-truth 3D geometry, detection fell to near chance only when every valid query SIFT keypoint was given its ground-truth 3D coordinate. Within the evaluated RGB-only MDE attacks and warp models, the method serves as a software-layer defense that is light enough to run in a median of 474 ms on a consumer phone CPU. Designing stronger features for attackers who know the intrinsics remains future work.